# Distinct Amplitude Mode Dynamics Upon Resonant and Off-resonant Excitation Across the Charge Density Wave Energy Gap in $LaTe_3$ Investigated by Time- and Angle-resolved Photoemission Spectroscopy

Kecheng Liu[1], Takeshi Suzuki[1], Yigui Zhong[1], Teruto Kanai[1], Jiro Itatani[1], Linda Ye[2], Maya Martinez[2], Anisha Singh[2], Ian R. Fisher[2], Uwe Bovensiepen[1,3], Kozo Okazaki[1]*

[1] *Institute for Solid State Physics, The University of Tokyo, Kashiwa, Chiba, 277-8581, Japan*
[2] *Geballe Laboratory for Advanced Materials and Department of Applied Physics, Stanford University, Stanford, CA 94305-4045, USA*
[3] *Faculty of Physics, University of Duisburg-Essen, 47048 Duisburg, Germany*

Non-equilibrium states generated by ultrafast laser pulses are characterized by specific phenomena that are not accessible in static measurements. Previous time- and angle-resolved photoemission spectroscopy (TARPES) studies on rare-earth tritelluride materials have revealed the laser-driven melting of the charge density wave order as well as its collective amplitude mode excitation. Variation of the excess energy deposited by optical pumping in the material promises pathways to control the dynamic material response. To this end, we use an optical parametric amplifier to generate a tunable pump photon energy. Studying $LaTe_3$ we compare the dynamics driven by pumping resonantly across the charge density wave energy gap with the effect of pumping at a twice higher photon energy in a TARPES pump-probe experiment. We clearly identify a pump photon energy dependent behavior. At the larger pump photon energy, the excess electronic energy generates lattice heating mediated by e-ph coupling and softening of the amplitude mode frequency from 3 to 2 THz. Remarkably, the resonant pumping across the CDW gap results in a time-independent amplitude mode frequency. We conclude that the resonant excitation across the energy gap excites the amplitude mode selectively while additional electronic excess energy deposited at higher pump photon energy modifies the crystal properties transiently by incoherent dissipative processes.

## 1. Introduction

The charge density wave (CDW) is a broken symmetry ground state in low-dimensional materials that emerges below a transition temperature $T_{CDW}$, where the electron density is spatially modulated with a periodicity distinct from the high temperature state. This phenomenon occurs due to the spontaneous lattice distortion and is linked to energy gain which results in opening of a gap in the Fermi surface[1]. Although CDW formation can be described by effective mean field models in thermal equilibrium, it is challenging to distinguish the charge and lattice contributions due to the strong electron-phonon coupling. Pump-probe experiments using femtosecond laser pulses offer, in contrast to static measurements, opportunities to detect the electron and lattice subsystems separately by exciting the system into non-equilibrium states and probe the laser-pump induced dynamics in the time domain [2,3,4]

Rare-earth tritellurides ($R$Te$_3$, $R$ is a rare-earth element) possess a layered and weakly orthorhombic structure and are known as a class of CDW materials[5] in which the ultrafast dynamics have been studied widely [4,6,7,8,9,10,11]. In the normal-state, the Fermi surface originates from the nearly square-shaped Te sheets. The previous results of angle-resolved photoemission spectroscopy (ARPES) have revealed an imperfect Fermi surface nesting along the $c$ axis, where the vector $\mathbf{q}_{CDW}$ creates a CDW gap $2\Delta$, leading to an incomplete diamond-shaped Fermi surface. The contraction of rare-earth ions with increasing atomic number $Z$ in $R$Te layers induces increasing chemical pressure which results in a decreasing $T_{CDW}$ and gap size[12]. Besides, $R$Te$_3$ materials with heavier $R$ atoms (Tb-Tm) exhibit a second CDW transition at lower temperature along the $a$ axis [5].

Time- and angle-resolved photoemission spectroscopy (TARPES) [4,13,14,15,16] is widely applied to probe the dynamics of electronic order of the CDW with femtosecond laser pulses. For example, the suppression of $2\Delta$ can be seen as a series of energy-momentum cuts with varying delay time. Such pump-probe spectroscopy studies have demonstrated the narrowing process in the CDW gap and the pump-fluence-dependent CDW amplitude mode (AM). Experimental and theoretical studies revealed how the CDW gap evolves after the photoexcitation with a pump fluence dependent response of the CDW AM [4,9,10,17]. The frequency of the amplitude mode observed in these experiments allows to analyze the transient potential energy surface upon excitation. Under sufficiently strong excitation that modifies or even melts the CDW the potential energy surface changes transiently from a double well to a single well potential, and back to double well potential upon excess energy relaxation. This phenomenon results in a delayed softening of the amplitude mode as reported [9,10]. Typically, pump-probe experiments generate the non-equilibrium state by absorbing near-infrared laser pulses at a photon energy of 1.5 eV as the fundamental of Ti:sapphire laser sources. Several investigations have demonstrated that variation of the pump photon energy can change the dynamic phenomena and provide insights not accessible with pumping at 1.5 eV, for example, Refs. 18,19,20,21,22,23,24. It was discussed specifically for $DyTe_3$ that the electronic excess energy could influence the competing trends of pump-induced gap melting and enhanced Fermi surface nesting [17)].

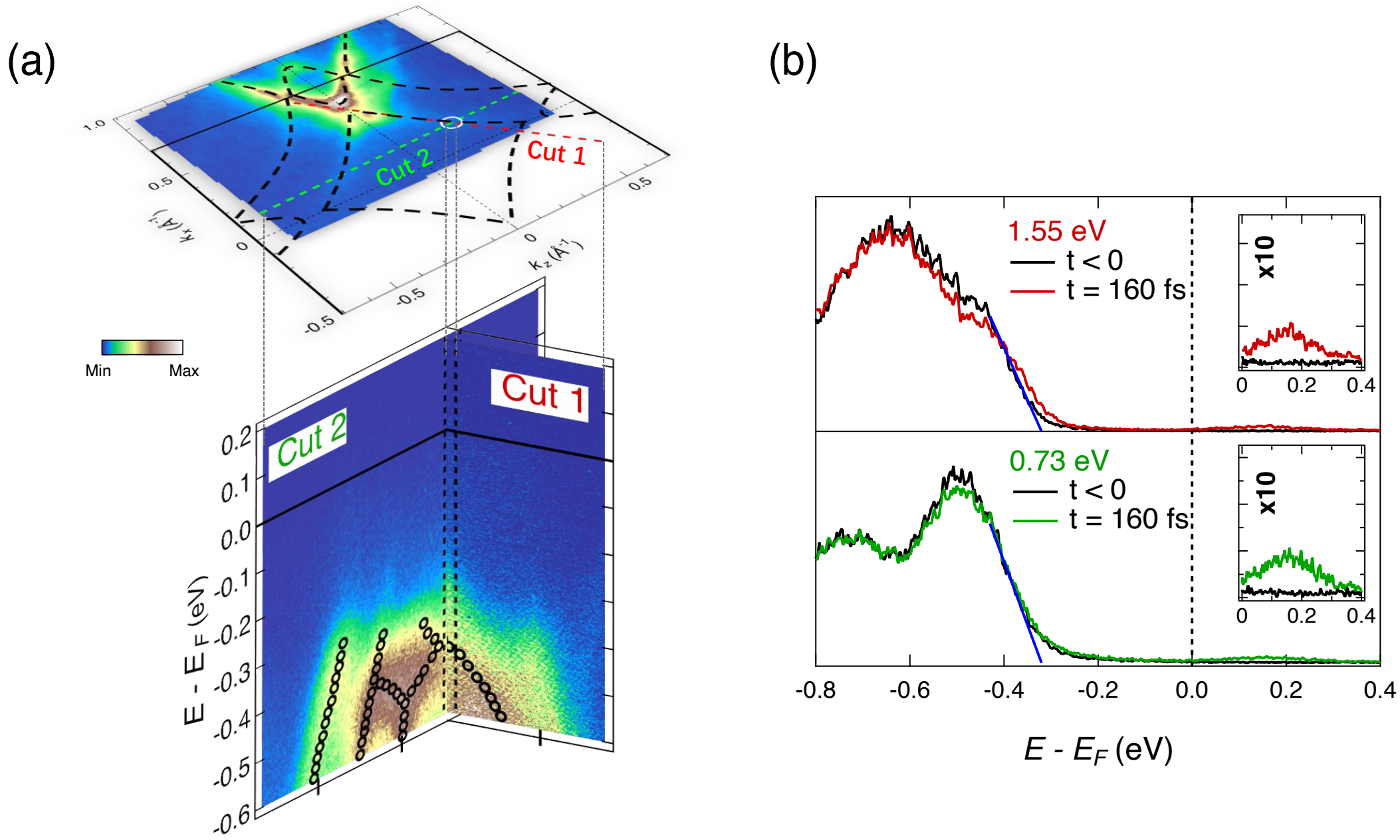

Fig. 1. (Color online) (a) Static ARPES data of $LaTe_3$ using a Helium discharge lamp (*hv* = 21.2 eV) at room temperature. The upper panel shows the FS map of $LaTe_3$ where the solid line corresponds to the first Brillouin zone and the dashed lines mark the continuous FS without CDW modulation. The red and green dashed lines show the momentum cut of the TARPES measurements for pumping at 1.55 eV and 0.73 eV photon energy, respectively. The region around the intersection of the two cuts is marked by the white circle. The lower panel gives the *E-k* intensity maps of cut 1 and cut 2, where the black markers indicate the band dispersion. The dashed area corresponds to the CDW gap region by the circle marked in the FS map. (b) EDCs recorded with HHG light at *hv* = 21.7 eV before (black) and after (colored) the pump arrival, with the momentum integrated in the region between the dashed lines in (a). Blue lines indicate the leading edges of the occupied-state bands. The insets present the EDCs for $E > E_F$ with the intensity multiplied by a factor of 10.

In this TARPES work we compare the dynamics driven by a near-infrared (NIR) pump pulse at a photon energy $hv^{NIR}_{pump}$=1.55 eV used in previous TARPES studies of $R$Te$_3$ with the dynamics triggered by $hv^{MIR}_{pump} = 2\Delta$ which is in the mid-infrared (MIR) spectral range. We observe a pronounced difference in the frequency at which the amplitude mode responds to the pump excitation. At NIR pumping the amplitude mode frequency softens from 3 to 2 THz after 1 ps in agreement with earlier reports [9,10]. More importantly, for MIR pump, a single frequency amplitude mode which oscillates at a constant frequency of 3 THz is identified. This difference is associated with the excess energy deposited in the material by the higher pump photon energy $h\nu^{NIR}{}_{pump} > h\nu^{MIR}{}_{pump}$.

## 2. Experimental

The real part of the optical conductivity $\sigma_1$, which is proportional to $\omega\varepsilon_2$, where $\varepsilon_2$ is the imaginary part of the dielectric function and ω is the optical frequency, allows an estimation of the resonant absorption across 2Δ [25]. For $LaTe_3$ which is the $R$$Te_3$ material with the largest in-plane lattice constant, largest energy gap, and largest gapped fraction of the Fermi surface [12], we estimate the optical absorption maximum to 0.7 eV [26,27]. The lower pump photon energy $h\nu^{MIR}{}_{pump}$ = 0.7 eV matches 2Δ and the larger one $h\nu^{NIR}{}_{pump}$ = 1.5 eV deposits additional electron energy at higher energies in the band. The incident pump fluence $F$ was determined to 1.0±0.3 mJ/cm$^2$ by dividing the incoming energy per pulse by the illuminated surface area of the sample, which was adjusted to keep $F$ constant for $h\nu^{NIR}{}_{pump}$ and $h\nu^{MIR}{}_{pump}$. For a comparison of the absorbed fluences at these two pump photon energies we estimate the ratio of the corresponding absorption coefficients $\alpha$(0.7 eV) / $\alpha$(1.5 eV). Since $\alpha = \omega\varepsilon_2/c$, $c$ is the velocity of light in vacuum [28], $\alpha$(0.7 eV) / $\alpha$(1.5 eV) = $\sigma_1$(0.7 eV) / $\sigma_1$(1.5 eV). Taking data reported in Refs. 25, 26, 27 into account we find that $\alpha$(0.7 eV) / $\alpha$(1.5 eV) varies between 2 and 3 at temperatures far below $T_{CDW}$. Therefore, we conclude that the absorbed fluence at $h\nu^{MIR}{}_{pump}$ is by that factor larger than at $h\nu^{NIR}{}_{pump}$. Absorbed pump fluence for NIR and MIR are therefore in the regime of strong excitation fluences that modify the potential energy surface from a doubled well to a single well potential and back [9,10].

The experimental setup used for this TARPES study was described recently [29]. It is equipped with a gas phase high harmonic generation beamline providing femtosecond pulses at 21.7 eV probe photon energy combined with a tunable pump photon energy using an optical parametric amplifier, a He discharge lamp for static ARPES, and a vacuum chamber with a photoelectron spectrometer. For static ARPES the energy resolution is 15 meV and for TARPES using the high harmonic light the energy resolution is 200 meV. The temporal resolution is 50 fs for NIR pumping and 60 fs for MIR pumping [29]. The repetition rate of the TARPES experiment is 10 kHz. The samples are cooled by liquid helium cryostat. TARPES experiments were conducted at 30 K sample temperature with the cleaved sample kept in ultrahigh vacuum. In the experiments we determined the Fermi energy $E_F$ by a reference measurement on a gold surface.

## 3. Results

The dynamics driven by femtosecond laser pulses are investigated by time- and angle-resolved photoelectron. We carried out the static ARPES measurements to analyze the electronic structure in thermal equilibrium at $T$ = 300 K, which is well below the CDW transition temperature of $LaTe_3$ estimated to 670 K [27)]. The upper panel of Fig. 1(a) shows the Fermi surface (FS) map of $LaTe_3$ integrated from -10 meV to 10 meV across $E_F$, where the diamond-shaped FS sheet originates from the overlapping $5p_x$ and $5p_z$ orbitals in the planes of Te atoms. Absence of the intensity at $E_F$ results from the opening of the CDW gap in regions of the Brillouin zone with a well nested FS. To study the CDW dynamics, we selected two different momentum cuts containing the CDW signature, shown as the green and red dashed lines in FS map for the TARPES measurements presented below. Cut 1 provides information along the presumed high temperature FS while cut 2 crosses the CDW-gapped area to a second branch of the FS. The momentum- and energy-dependent photoelectron intensity as static $E$-$k$ maps which represents the electronic structure is shown in the lower panel of Fig. 1(a). The investigation of these two different cuts is the consequence of (i) limitations to determine prior to ARPES analysis the in-plane crystal direction $x$ along which the CDW modulates the electron density versus $z$ along which no modulation is found, (ii) sensitivity of the samples to ambient conditions which prevents determination of the crystal directions using Laue diffraction, and (iii) absence of an azimuthal degree of freedom in crystal rotation in the ARPES instrument. As a result, when mounting the samples based on the long edge of the as-grown rectangular crystals, there is an approximately 50% probability that the in-plane orientation corresponds to either the [100] direction (resulting in cut 2) or the [101] direction (resulting in cut 1), leading to different momentum cuts being probed in different measurements.

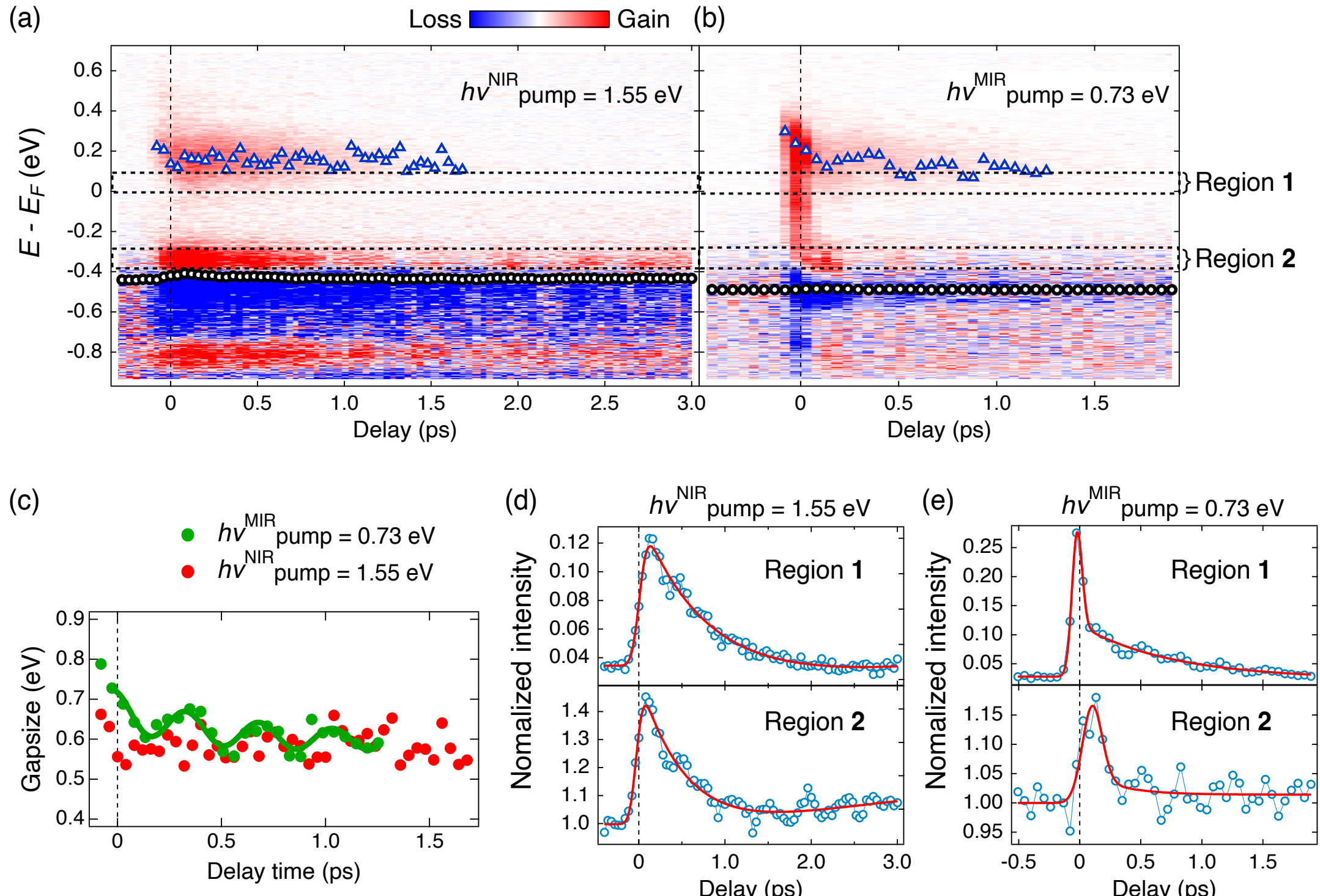

Fig. 2. (Color online) (a, b) Temporal evolution of the differential EDCs in a color map at $k_x$=0.17 Å$^{-1}$, and $k_z$=0.25 Å$^{-1}$, integrated within the area indicated by the circle in the upper panel of Fig. 1(a), pumped by 1.55 eV (a) and 0.73 eV (b), respectively. The blue and black markers indicate the band maxima above and below the CDW gap. (c) Time-dependent change of the CDW gap for $h\nu^{NIR}{}_{pump}$ and $h\nu^{MIR}{}_{pump}$. (d, e) The time-dependent intensity integrated in the energy regions marked by the black dashed boxes in panels (a) and (b), labeled by "Region 1" and "Region 2". The intensities are normalized by the average intensity for $t$<0 in "Region 2", for each pump energy. The red lines depict fits detailed in the main text.

In the following, we investigate the nonequilibrium properties in the nested region of the FS by TARPES upon femtosecond laser excitation. Figure 1(b) shows the energy distribution curves (EDCs) before and after the excitation by the pump pulse, integrated in the region of the Brillouin zone indicated by the white circle in FS map. To confirm that the two EDCs in Fig. 1(b) correspond to the same momentum position (*i.e.*, the intersection point of cut 1 and cut 2), we added blue lines indicating the leading edges of the occupied-state bands. These lines intersect the energy axis at $E$-$E_F$ = -0.32 eV for both EDCs, confirming that they were taken at equivalent momenta. Although the spectral shapes of the two EDCs appear different, this arises from the ~45° difference in azimuthal angle of the sample with respect to the polarization of the probe pulse between cut 1 and cut 2. This difference modifies the photoemission matrix elements and thus the spectral weight. However, it does not affect the leading-edge position of the spectra.

Figure 2(a,b) shows the temporal evolution of the differential EDCs as a function of pump-probe time delay $t$ for $h\nu^{NIR}{}_{pump}$ and $h\nu^{MIR}{}_{pump}$. The red and blue colors indicate the pump-induced gain and loss in intensity, respectively, with respect to the EDC before pumping at $t$ < 0. For the 1.55 eV pump, as shown in Fig. 2(a), the increasing intensity between $E$-$E_F$ = -0.4

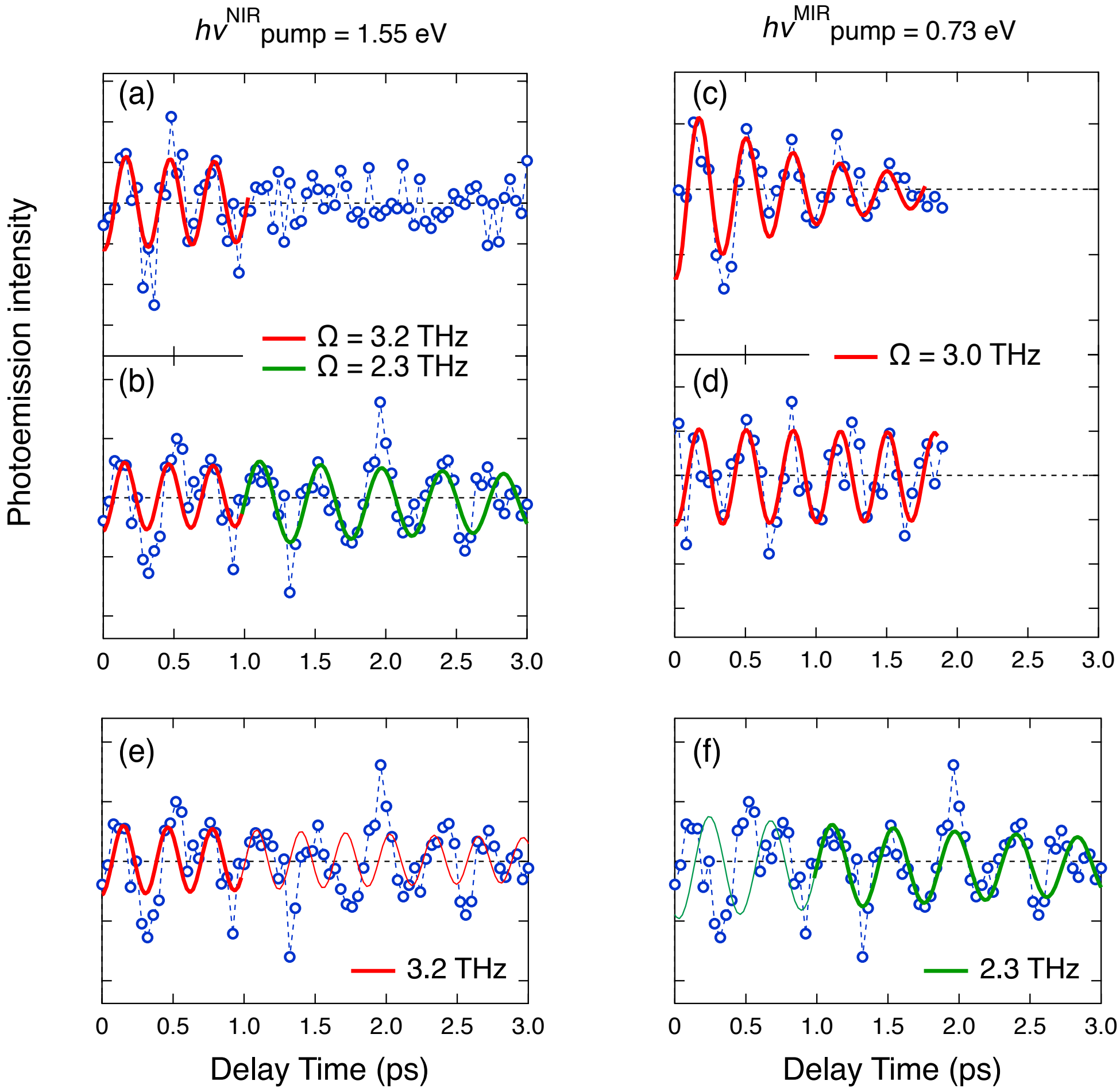

Fig. 3. (Color online) (a-d) The residual photoemission intensity as a function of delay time after subtraction of the non-oscillatory component. The colored solid lines are the fitting results of a cosine function multiplied with an exponential function. Panels (a,c) represent intensity changes obtained in the energy interval above $E_F$ (Region 1 in Fig. 2), panels (b,d) show data obtained in the energy interval below $E_F$ (Region 2 in Fig. 2). The time intervals for fitting of the higher frequency (red line) is from 0.00 to 1.04 ps, and for the lower frequency (green line) is from 1 to 2.5 ps. This results in error bars in the frequency determination of 3.20±0.15 THz for the higher frequency and 2.3±0.2 THz for the lower frequency. Panels (e,f) show fits of the data in panel (b) with single frequencies of 3.2 THz and 2.3 THz, respectively.

and -0.25 eV represents a shift of the band edge towards higher energy. In combination with the energy of the band edge 0.2 eV above $E_F$ we conclude to observe a reduction of the gap size upon pumping. For the 0.73 eV pump, see Fig. 2(b), the intensity increases between $E$-$E_F$ = -0.2 eV and $E_F$ during the overlap of pump and probe. This effect is attributed to a replica band created by the laser-assisted photoelectric effect (LAPE) [30]. The time this replica exists provides an estimate of the pump-probe cross correlation of 150 fs. At $t > 200$ fs periodic shifts in the band energies are recognized above and below the CDW gap. These oscillations are identified for several picoseconds and will be discussed in further detail below.

By analysis of the population of the upper as well as of the lower CDW gap edges, we determine the temporal evolution of the CDW gap quantitatively from the EDCs, see markers in Figs. 2(a,b). The energy of the state representing the upper limit of energy gap above $E_F$ is marked by blue triangles. We fit the band edge below $E_F$ with two Lorentzian functions and a linear background to represent the measured EDC. The obtained transient peak energies of the band edge below $E_F$ are depicted by black circles shown in Figs. 2(a,b). The transient CDW gap size is determined by the separation between the blue triangles and black circles. At $t$=0 we find values for $2\Delta$ of 0.65 eV for NIR pumping and 0.79 eV for MIR pumping. While the latter might overestimate $2\Delta$ due to LAPE, both values are larger than early reports that assumed a symmetric gap around $E_F$ [12], but they agree well with the results of optical spectroscopy [27]. The energy difference of the band edges below and above $E_F$ as a function of pump-probe time delay provide the temporal evolution of $2\Delta$ for both pump energies. This result is shown in Fig. 2(c). For $h\nu^{NIR}{}_{pump}$ and $h\nu^{MIR}{}_{pump}$ the energy gap is barely reduced. Due to existence of the replica band in the photoemission spectra for pumping by MIR photons, the value of $2\Delta$ observed initially appears to slightly differ from pumping with NIR photons. At $t > 0.5$ ps the gap size is consistently found for both pump photon energies to be $2\Delta = 0.60\pm0.05$ eV. For MIR pumping we resolve oscillations of $2\Delta$ with a frequency of $2.91 \pm 0.12$ THz as determined by the fit depicted in Fig. 2(c). These gap oscillations represent the amplitude mode [1] and the frequency of ~3 THz agrees with previous reports [11,17]. Fig. 2(c) also shows that upon pumping at 1.55 eV the amplitude mode was not resolved contrary to previous reports [11,17] which we assign to error bars obtained in fitting the upper CDW band edge.

To gain more insight, we integrate the intensity in the energy intervals indicated by the black dashed lines in Figs. 2(a,b). The resulting integral intensities are shown in Figs. 2(d,e) as a function of $t$. The center of the energy interval is tuned to the band edges above and below the CDW gap to track the pump-induced oscillations in the photoemission intensity. For both, $h\nu^{NIR}_{pump}$ and $h\nu^{MIR}_{pump}$, the time-dependent intensities are normalized by the energy integrated intensity within region 2 at $t<0$. To determine the non-oscillatory component in the transient data we fitted the time-dependent integrated intensity with a double exponential function, as indicated by the red lines in Figs. 2(d,e). In the fitting of region 1 for $h\nu^{MIR}_{pump}$, an additional Gaussian component centered at $t_0$ was included to account for the LAPE-induced peak. This improves the fitting around $t_0$ without affecting the extracted oscillatory component. In the lower panel of Fig. 2(d), an increase of the intensity is noticed for pumping at $h\nu^{NIR}_{pump} = 1.55$ eV for $t>1.5$ ps. Such an increase is not observed for pumping at $h\nu^{MIR}_{pump} = 0.73$ eV. We come back to this observation in the discussion section below. By subtracting the non-oscillatory component from the time-dependent integrated intensity, we obtain the oscillatory response detected in TARPES as shown in Figs. 3(a-d). We use an oscillatory function with a single frequency Ω to fit the residual intensity in certain time intervals as marked by the colored solid lines. In Fig. 3(a), for the 1.55 eV pump, the oscillation in the band above $E_F$ becomes unrecognizable after 1.0 ps. For the band edge below $E_F$, the oscillation frequency is determined at early delays $t < 1.0$ ps to $\Omega = 3.2$ THz, see the red line. This frequency is identical to the one obtained above $E_F$. Once the oscillations above $E_F$ have disappeared with increasing $t$, the frequency below $E_F$ drops $t > 1.0$ ps to $\Omega = 2.3$ THz, see the green line. Fitting of the NIR pumped intensity oscillations with a single frequency of either 2.3 THz or 3.2 THz does not lead to successful description of the data as illustrated in Fig. 3 (e,f). To verify our fitting analysis, we perform fast Fourier transformation (FFT) of the oscillatory photoemission intensity variations. The corresponding results are shown in Fig. 4. All the peak positions of the FFT spectra agree well with the frequency obtained by the fitting procedure. Note that FFT was performed separately for the time intervals $t > 1.0$ ps and $t < 1.0$ ps and are plotted by the green and red lines in Fig. 4(b), respectively.

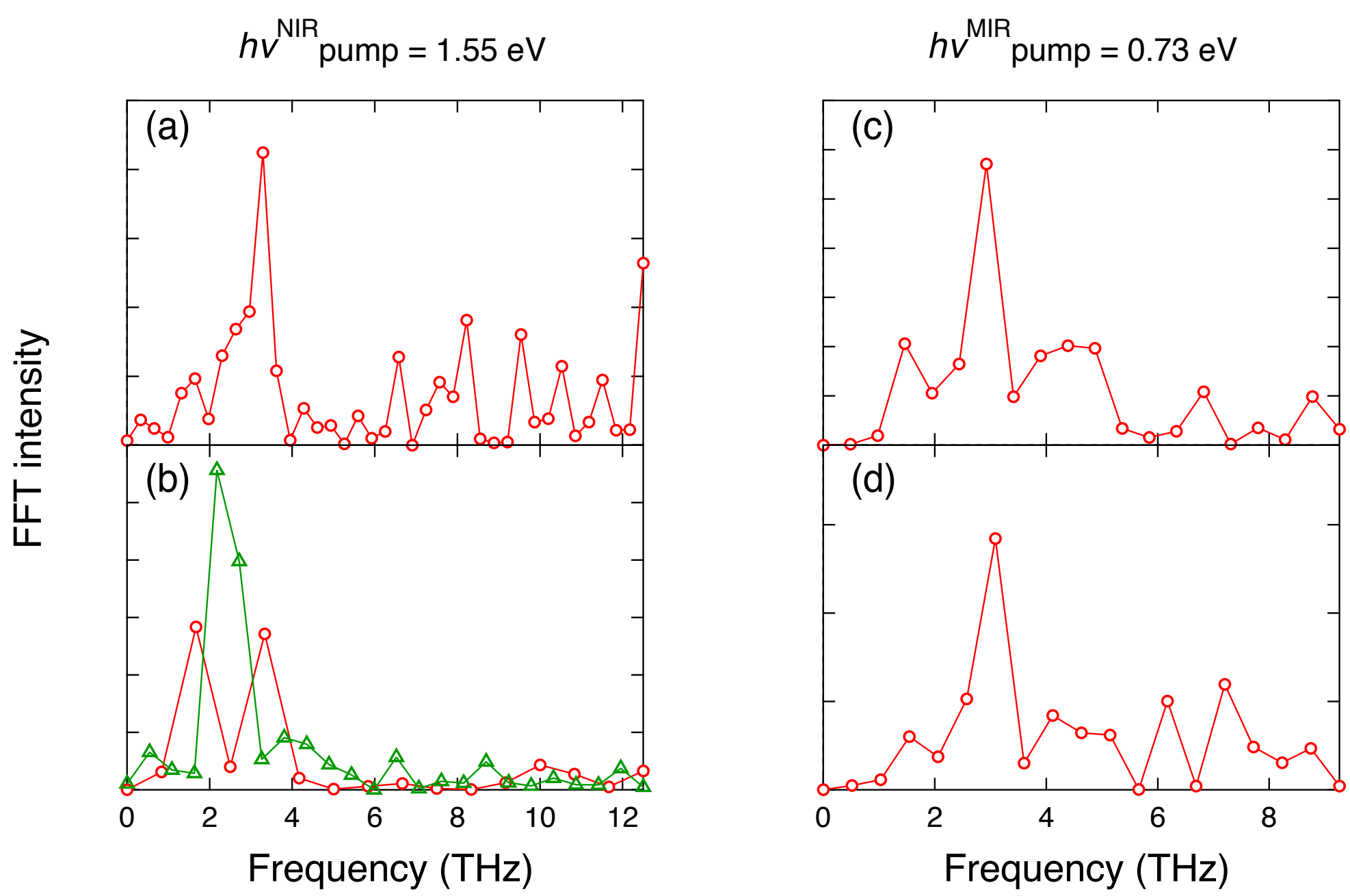

Fig. 4. (Color online) Panels (a-d) present the fast Fourier transformation of the data shown in Figure 3. Panels (a,b) depict data for $h\nu^{NIR}_{pump}$=1.55 eV and panels (c,d) show data for $h\nu^{MIR}_{pump}$=0.73 eV. The red (green) spectrum in panel (b) represent the time-dependent data in Fig. 3(b) for $0<t<1$ ps ($1<t<3$ ps).

**4. Discussion**

The observation of different oscillatory response to pumping by $h\nu^{NIR}_{pump}$ and $h\nu^{MIR}_{pump}$ requires further attention. The time-dependent $2\Delta$, as shown in Fig. 2(c), is dominated by the transient energy of the CDW band edge above $E_F$ since the energy of the band edge below the $E_F$ (black markers) varies weakly for both pump photon energies. For the 0.73 eV pump, which is resonant with $2\Delta$, the position of the CDW band edge above $E_F$ exhibits a well observed oscillatory behavior as a function of $t$. This is the reason for the observed amplitude mode oscillations in $2\Delta$ upon pumping at $h\nu^{MIR}_{pump}$, as shown in Fig. 2(c). Besides the transient gap size variations, the time-dependent photoemission intensity also exhibits a dependence on the pump photon energy. For the case of 1.55 eV pump photon energy, a continuous increase of the band edge intensity below the CDW gap it is recognized for $t > 1.5$ ps in region 2, see Fig. 2(d). For MIR pumping at 0.73 eV photon energy, electrons are excited to final states just above the CDW energy gap into the CDW band minimum while electrons excited by the NIR pump at 1.55 eV occupy final states higher up in the band. The corresponding excess energy for the latter excitation is expected to be dissipated through secondary excitations of the material, mediated by e–e and e–ph scattering. Due to large CDW energy gap and the large gapped fraction of the Fermi surface, see Fig. 1(a) and Ref. 12, e-e scattering is likely less important due to phase space constraints [31] than e-ph coupling which leads to lattice heating.

Such lattice heating has been observed in time-resolved electron diffraction experiments of $LaTe_3$ by an increase in thermal diffuse scattering upon NIR pumping at time delays of 1 ps [6,32] supporting our assignment of the TARPES intensity increase observed at $t > 1$ ps.

The dynamics of the coherent phonon oscillations upon NIR pumping at 1.55 eV changes after $t = 1.0$ ps. This time delay is close to the one at which photoemission intensity starts to increase again after the early intensity drop, see Fig. 2(d). Prior to 1 ps, the 1.55 eV pump, similarly to the 0.73 eV case, excites an oscillation with a frequency of 3.2 THz at both band edges, above and below $E_F$, as shown in Fig. 3(b). Since the static sample temperature and the incident pump fluence are identical for experiments using both pump photon energies, excitation of a similar dynamics by the two different pump pulses would not be surprising. As mentioned in section 2, the absorbed fluence at $h\nu^{MIR}{}_{pump}$ is larger the one at $h\nu^{NIR}{}_{pump}$ due to a larger absorption coefficient due to the across gap resonance. However, in case of pumping at 1.55 eV, the oscillation frequency changes from 3.2 to 2.3 THz at $t = 1.0$ ps, as monitored below $E_F$ in Fig. 3(b) in agreement with previous reports of such amplitude mode softening due to the pump-induced change of the potential energy surface of the CDW [9,10]. As an explanation we provide the following scenario. The NIR pumping, which deposits excess energy in the electronic system, induces a change in the crystal lattice due to dissipation of this excess energy. The time delay of 1 ps, at which the frequency change is found, agrees with typical lattice response times of metals in general [33] and $LaTe_3$ in particular [6,32]. Moreover, this time delay matches those concluded from simulated order-parameter dynamics using time-dependent Ginzburg-Landau simulations and three temperature model calculations to represent phonon and lattice excitations [9,10,11]. In such a transiently modified crystal lattice the frequency of the amplitude mode is likely to be different compared to a cold lattice at the equilibrium temperature. This scenario is based on the observation that almost simultaneously with the emergence of the 2.3 THz mode, the oscillation in the band edge above $E_F$ disappears, and the increase in intensity which we attribute to heating of the crystal lattice starts at this time delay for electronic states below $E_F$. The observed absence of the frequency change upon the resonant optical pumping across the gap at a photon energy $h\nu^{MIR}{}_{pump}$ is in good agreement with this explanation. Such resonant pumping avoids deposition of excess energy in the electronic system and limits energy dissipation into the crystal lattice although the absorption coefficient is due to the resonance larger than at the higher photon energy. We infer that in a cold lattice the amplitude mode frequency remains at the initially excited value.

## 5. Conclusion

The dynamics driven by femtosecond laser pulses are investigated by time- and angle-resolved photoelectron emission spectroscopy in the charge density wave material $LaTe_3$ for two different pump photon energies. One is the widely applied fundamental emission of the Ti:sapphire laser at 1.55 eV in the near infrared spectral region. The second one is the resonant absorption across the charge density wave gap at 0.7 eV in the mid infrared spectral region. We find notable differences in the dynamics of the coherently excited amplitude mode. Upon near infrared pumping the amplitude mode frequency softens from 3 to 2 THz at a time delay of 1 ps at which the crystal lattice is excited by e-ph coupling due to energy dissipation of the excess electronic energy deposited at the larger pump photon energy. In the case of the mid infrared pumping across the energy gap such electronic excess energy is minimized and the 3 THz amplitude mode frequency remains unchanged with increasing time delay. These results highlight the advantage of MIR excitation in resolving intrinsic amplitude mode dynamics even at high absorbed fluence, without inducing frequency softening. Our experimental results demonstrate the potential for controlling microscopic dynamic processes by the chosen pump photon energy and are expected to motivate further investigation in the future.

**Acknowledgment**

This work was supported by Grants-in-Aid for Scientific Research (KAKENHI) (Grant Nos. JP19H01818, JP19H00659, JP19H00651, JP24K01375, JP24K00565, and JP24KF0021) from the Japan Society for the Promotion of Science (JSPS), by JSPS KAKENHI on Innovative Areas "Quantum Liquid Crystals" (Grant No. JP19H05826), and the Quantum Leap Flagship Program (Q-LEAP) (Grant No. JPMXS0118068681) from the Ministry of Education, Culture, Sports, Science, and Technology, Japan (MEXT). Crystal growth and characterization at Stanford was supported by the Department of Energy, Office of Basic Energy Sciences, under contract DE-AC02-76SF00515. Funding by the Deutsche Forschungsgemeinschaft (DFG, German Research Foundation) within Project ID No. 278162697-SFB 1242 and through Project No. BO1823/12-FOR 5249 (QUAST) is gratefully acknowledged. U.B. is grateful for the hospitality of the Institute for Solid State Physics, University of Tokyo, and acknowledges support through the visiting professor program of the institute.

*E-mail: okazaki@issp.u-tokyo.ac.jp